\documentclass[prl,amsmath,amssymb,twocolumn,showpacs,floatfix]{revtex4}
\usepackage{graphicx}
\usepackage{dcolumn}
\usepackage{bm}

\begin{document}

\title{Detection of the Unusual Magnetic Orders in the Pseudogap Region of a 
High-Temperature Superconducting YBa$_2$Cu$_3$O$_{6.6}$ Crystal by Muon-Spin
Relaxation}

\author{J.E.~Sonier,$^1$ V.~Pacradouni,$^1$ S.A. Sabok-Sayr,$^1$ W.N.~Hardy,$^2$ D.A.~Bonn$^2$, 
R.~Liang$^2$ and H.A.~Mook$^3$}

\affiliation{$^1$Department of Physics, Simon Fraser University, Burnaby, British Columbia V5A 1S6, Canada \\
$^2$Department of Physics and Astronomy, University of British Columbia, Vancouver, British 
Columbia V6T 1Z1, Canada\\ 
$^3$Oak Ridge National Laboratory, Oak Ridge, Tennessee 37831, USA}

\date{\today}

\begin{abstract}
We present muon spin relaxation ($\mu$SR) measurements on a large 
YBa$_2$Cu$_3$O$_{6.6}$ single crystal in which two kinds of unusual magnetic order have 
been detected in the pseudogap region by neutron scattering. A comparison is made
to measurements on smaller, higher quality YBa$_2$Cu$_3$O$_y$ single crystals. 
One type of magnetic order is observed in all samples, but does not evolve significantly
with hole doping. A second type of unusual magnetic order is observed only in the 
YBa$_2$Cu$_3$O$_{6.6}$ single crystal. This magnetism has an ordered magnetic moment that 
is quantitatively consistent with the neutron experiments, but is confined to just a small 
volume of the sample ($\sim \! 3$~\%). Our findings do not support theories that 
ascribe the pseudogap to a state characterized by loop-current order, but instead 
indicate that dilute impurity phases are the source of the unusual magnetic orders
in YBa$_2$Cu$_3$O$_y$.
\end{abstract}

\maketitle
It is widely believed that the mysterious pseudogap region of high-transition temperature (high-$T_c$) 
copper oxide superconductors is caused by a ``hidden order''. 
Varma \cite{Varma:97,Varma:06} has proposed that the pseudogap is caused by a 
circulating-current (CC) state that breaks time-reversal symmetry and is characterized by a unique 
long-range pattern of loop currents in the CuO$_2$ planes that breaks rotational symmetry, but preserves 
the translational symmetry of the lattice (TSL). Alternatively, Chakravarty {\it et al.} 
\cite{Chakravarty:01} has attributed the pseudogap to a competing $d$-density wave (DDW) order.
The DDW phase also breaks time reversal and rotational symmetries, but has an orbital 
current pattern that breaks the TSL. In both models the loop-current order is predicted 
to weaken with increased doping, and to vanish at a quantum critical point
somewhat above optimal doping.

Seemingly direct evidence for DDW order comes from neutron scattering 
experiments on underdoped YBa$_2$Cu$_3$O$_y$ (YBCO) with $y$ \! = \! 6.6 \cite{Mook:01,Mook:02} 
and $y$ \! = \! 6.45 
\cite{Mook:04}, which reveal a weak antiferromagnetic (AF) ordered magnetic moment predominantly 
directed perpendicular to the CuO$_2$ planes. However, no such static magnetic order was observed
by Stock {\it et al.} \cite{Stock:02} in a neutron study of YBa$_2$Cu$_3$O$_{6.5}$.     
A second unusual magnetic order recently detected in the 
pseudogap region of YBCO and HgBa$_2$CuO$_{4+\delta}$ (Hg1201)
by polarized neutron scattering \cite{Fauque:06,Mook:08,Li:08} does not break the TSL, 
and hence is instead qualitatively consistent with the CC phase. However, the ordered moment in this 
case is not perpendicular to the CuO$_2$ planes as expected for the CC phase, but rather has a large 
in-plane component. Spin-orbit coupling \cite{Aji:07} or orbital currents involving the apical 
oxygens \cite{Weber:09} have been offered as possible reasons for why the magnetic moments are severely 
canted. The first scenario has also been invoked \cite{Aji:08} to explain the onset of an accompanying 
weak ferromagnetism near $T^*$, which has been detected in YBCO by high-resolution polar Kerr effect 
(PKE) experiments \cite{Xia:08}. 

What is most surprising is the lack of evidence from previous zero-field (ZF) $\mu$SR experiments  
for the existence of loop-current magnetic order. While fields on the order of 100~G are expected 
in a $\mu$SR experiment, no field of this size has been detected in YBCO \cite{Sonier:01,Sonier:02} or 
La$_{2-x}$Sr$_x$CuO$_4$ \cite{MacDougall:08}. To reconcile this discrepancy it has been suggested that 
charge screening of the positively charged muon ($\mu^+$) severely underdopes its local environment, 
causing the loop-current-order to vanish over a distance of several lattice constants \cite{Shekhter:08}.
However, the assertion that the muon strongly perturbs its local magnetic environment runs contrary
to the excellent agreement between $\mu$SR and nuclear magnetic resonance (NMR) measurements of
AF correlations in underdoped cuprates \cite{Julien:03}. Furthermore, we demonstrate here that weak 
magnetic fields detected in YBCO are not caused by undisturbed loop-current-order located an 
appreciable distance from the muon. 

\begin{figure}
\centering
\includegraphics[width=9.0cm]{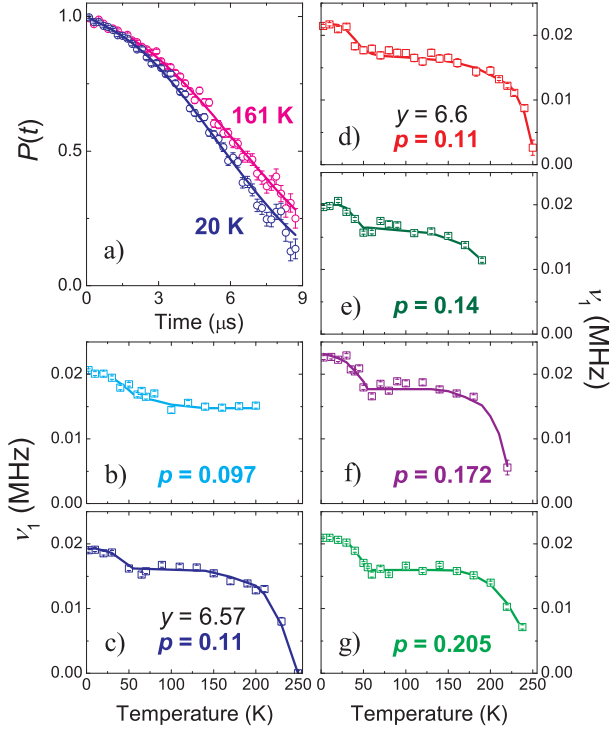}
\caption{(Color online) (a) Representative ZF-$\mu$SR signals for the large $y \! = \! 6.6$ 
single crystal. The solid curves are fits to Eq.~(\ref{eq:pol}). 
Temperature dependence of $\nu_1$ for the smaller YBCO single crystals with 
oxygen content (b) $y \! = \! 6.50$, (c) $y \! = \! 6.57$, (e) $y \! = \! 6.80$, (f) $y \! = \! 6.95$, 
for (g) the small Ca-doped YBCO single crystals with $y \! = \! 6.98$, and (d) the large 
$y \! = \! 6.6$ single crystal. 
The solid curves in (b)-(g) are guides to the eye. The error bars come from the fits. 
The hole-doping concentration $p$ for each sample is shown in the respective panel.}
\label{fig1}
\end{figure}

In contrast to 
neutron scattering, ZF-$\mu$SR is a local probe of magnetism that can distinguish magnetic and non-magnetic 
regions of the sample. The time evolution of the muon-spin polarization 
$P(t)$ for an ensemble of muons implanted one-by-one in the sample is dependent on the local magnetic 
field distribution, and is measured by detecting the muon-decay positrons. Magnetic order results in 
an oscillation of $P(t)$ with a frequency corresponding to the average field experienced by the muons. 
The ZF-$\mu$SR experiments reported here were performed on the M15 surface-muon channel at the 
Tri-University Meson 
Facility (TRIUMF), with the initial spin polarization $P(0)$ parallel to the CuO$_2$ planes. One of the 
samples measured is the same large single crystal of YBCO with $y \! = \! 6.6$ ($T_c \! = \! 62.7$~K) 
in which neutron experiments \cite{Mook:08,Mook:01,Mook:02} have detected 
both kinds of unusual magnetic orders in the pseudogap region. The $y \! = \! 6.6$ crystal has 
a cylindrical shape, with a diameter of 22.9~mm and a thickness of 10.8~mm. Measurements were also 
carried out on YBCO single crystals a thousandth of the size, grown by a self-flux method in
fabricated BaZrO$_3$ crucibles \cite{Liang:98}. Mosaics consisting 
of $\sim \! 10$ single crystals of thickness on the order of 0.1~mm, making up 
a total $a$-$b$ surface area of 20 to 30~mm$^2$, were measured for
$y$ \! = \! 6.50 ($T_c \! = \! 59$~K), $y \! = \! 6.57$ ($T_c \! = \! 62.5$~K), $y \! = \! 6.80$ 
($T_c \! = \! 84.5$~K), and $y \! = \! 6.95$ ($T_c \! = \! 93.2$~K), and for
Ca-doped YBCO with $y \! = \! 6.98$ ($T_c \! = \! 75$~K).

\begin{figure}
\centering
\includegraphics[width=9.0cm]{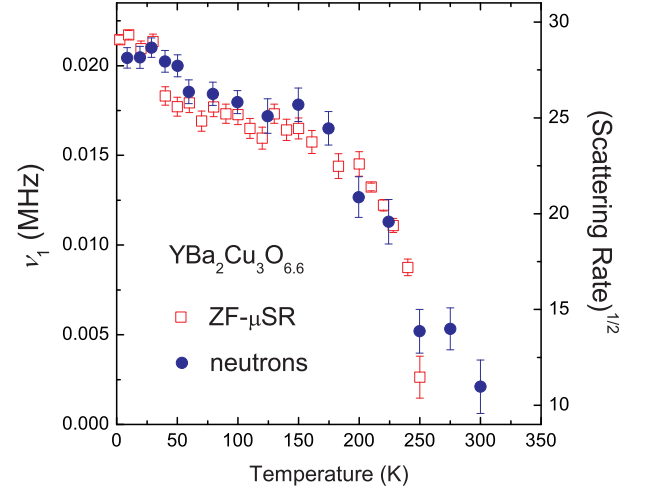}
\caption{(Color online) 
A comparison of the temperature dependence of $\nu_1$ for the large $y \! = \! 6.6$ single crystal 
with the square root of the neutron scattering rate at wavevector (0.5, 0.5, 2) from Ref.~\cite{Mook:01}. 
Both quantities are proportional to the ordered magnetic moment.}
\label{fig2}
\end{figure}

Figure~\ref{fig1}(a) shows typical ZF-$\mu$SR spectra for the large
$y \! = \! 6.6$ crystal. 
The spectra for all samples are well described over a 9~$\mu$s time range by the following
polarization function
\begin{eqnarray}
P(t) & = & P_{\rm KT}(t) \cos(2 \pi \nu_1 t) \nonumber \\
     & = & \left[\frac{1}{3}+\frac{2}{3}(1-\Delta^2 t^2) e^{-\Delta^2 t^2/2}\right] \cos(2 \pi \nu_1 t) \, ,
\label{eq:pol}
\end{eqnarray}
with a relaxation rate of $\Delta$ \! = 0.101264~$\mu$s$^{-1}$. The static Gaussian Kubo-Toyabe 
function $P_{\rm KT}(t)$ approximately describes the time evolution of the muon-spin polarization 
caused by the randomly oriented nuclear moments. The oscillatory component implies the presence of 
magnetic order that generates a field $B_1 \! = \! 2 \pi \nu_1/\gamma_{\mu}$ at the muon site, where
$\gamma_{\mu} \! = \! 0.0852$~$\mu$s$^{-1}$~G$^{-1}$ is the muon gyromagnetic ratio. Hence, 
the temperature dependence of the muon spin precession frequency $\nu_1$ (Figs~\ref{fig1}(b)-(g)) 
indicates changes in the average local field. 

Figure~\ref{fig2} shows $\nu_1$ for the $y \! = \! 6.6$ crystal superimposed on 
the temperature dependence of the square root of the neutron magnetic scattering \cite{Mook:01}. Both 
quantities are proportional to the size of the ordered moment. The striking agreement between these 
two data sets suggests that $\nu_1$ is caused by the same anomalous weak AF order detected by neutrons. 
In YBCO with $y \! = \! 6.6$, 
approximately 60~\% of the muons stop near a chain oxygen at (0.15, 0.44, 0.071) and 40~\% near an 
apical oxygen at (0.275, 0, 0.1333) \cite{Weber:90,Pinkpank:99}. Taking the ordered moment to be no 
larger than 0.02~$\mu_B$ (as established in Ref.~\cite{Mook:01}) and assuming DDW order 
\cite{Chakravarty:01}, the calculated dipolar field at the chain and apical oxygen muon sites is less 
than 5.2~G and 12.1~G, respectively. This is consistent with $B_1 \! \sim \! 1.6$~G at the lowest 
temperature. However, since the temperature dependence of $\nu_1$ is qualitatively similar 
in the other samples and evolves 
little with hole-doping concentration $p$ (see Figs~\ref{fig1}(b)-(g)), it is clear that the origin of the moments 
cannot be the orbital currents of the predicted DDW phase. Here we note that the onset of the 
pseudogap \cite{Ito:93} for YBCO with $y \! = \! 6.95$ ($p \! = \! 0.172$) is well within the 
temperature range of Fig.~\ref{fig1}(f). Furthermore, loop-current magnetic order is not expected to occur in 
the Ca-doped sample with $p \! = \! 0.205$. The tiny size of the ordered moment detected in 
Refs.~\cite{Mook:01,Mook:02,Mook:04} may be an indication that the magnetic order resides in a small 
volume fraction of the sample. One cannot tell this from the neutron measurements. Unfortunately the ZF-$\mu$SR 
measurements are equally unrevealing, because $\nu_1$ is comparable in size to the dipolar 
fields generated by the nuclear moments. 

\begin{figure}
\centering
\includegraphics[width=9.0cm]{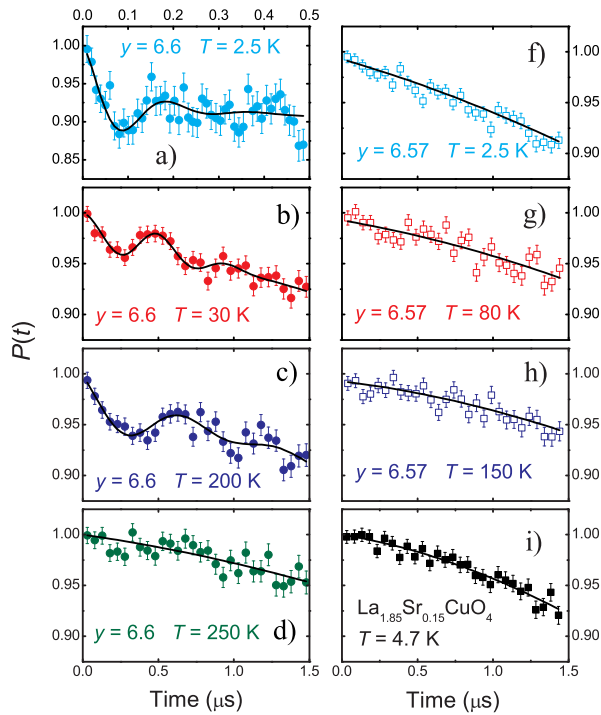}
\caption{(Color online) Time evolution of the ZF-$\mu$SR signal at early times for (a)-(d) the 
large $y \! = \! 6.6$ single crystal, (f)-(h) the mosaic of $y \! = \! 6.57$ single crystals, and
(i) a single crystal of La$_{1.85}$Sr$_{0.15}$CuO$_4$. The spectra are comprised of $\sim \! 11$ 
million muon-decay events, with the exception of (f) and (i) which have 3 and 5 times this number 
of statistics, respectively. The solid curve in each panel is a fit to Eq.~(\ref{eq:early}).
Note the difference in the time scale of (a) relative to (b)-(i).}
\label{fig3}
\end{figure}

\begin{figure}
\centering
\includegraphics[width=9.0cm]{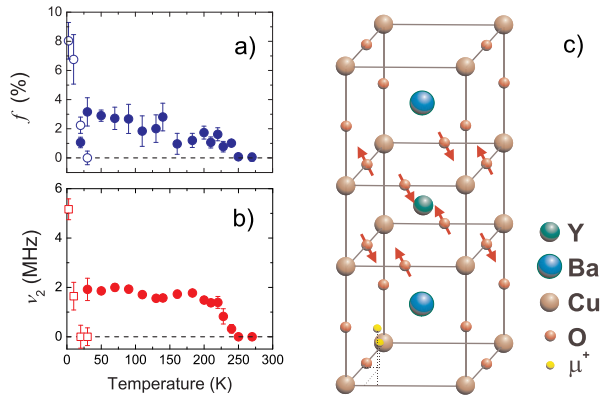}
\caption{(Color online)
Temperature dependence of (a) $f$ and (b) $\nu_2$ from fits of the ZF-$\mu$SR signals for 
the $y \! = \! 6.6$ single crystal to Eq.~(\ref{eq:early}). 
(c) A staggered arrangement of spins on the oxygen sites in the CuO$_2$ planes that accounts for 
the average field of the 3~\% oscillatory component. The spins are rotated in the CuO$_2$ 
planes $\pm~45^{\circ}$ from the $a$-axis, and are directed $55^{\circ}$ (or $235^{\circ}$) 
from the $c$-axis. Also shown are the two muon-stopping sites in YBCO with $y \! = \! 6.6$.}  
\label{fig4}
\end{figure}

Next we turn attention to the early times of the ZF-$\mu$SR spectra. Figures~\ref{fig3}(a)-(d) 
demonstrate a 
small-amplitude oscillatory signal below $T \! \sim \! 250$~K in the large $y \! = \! 6.6$ crystal. 
The ZF-$\mu$SR spectra over the first 1.5~$\mu$s are best described by the addition of a second 
oscillatory term to Eq.~(\ref{eq:pol})   
\begin{equation}
P(t) = (1-f)P_{\rm KT}(t) \cos(2 \pi \nu_1 t) + f e^{-\Lambda t} \cos(2 \pi \nu_2 t) \, ,
\label{eq:early}
\end{equation}
where $f$ is the volume fraction of the oscillating signal in Fig.~\ref{fig3}. The 
temperature dependence of $f$ and $\nu_2$ are shown in Fig.~\ref{fig4}. 
Below $T \! \sim \! 15$~K we observe a rapidly damped oscillatory signal coming from 
approximately 8~\% of the sample (see Fig.~\ref{fig3}(a)). At $T \! = \! 2.5$~K the oscillatory 
frequency is $\nu_2 \! = \! 5.1 \pm 0.4$~MHz, which corresponds to an average local field of 
$B_2 \! = \! 376 \pm 30$~G. We attribute this 8~\% component to the presence of the 
``green phase''  Y$_2$BaCuO$_5$ that is known to occur in this sample \cite{Mook:02} and undergo 
an AF transition at $T \! \sim \! 15$~K \cite{Kanoda:87}. Between $T \! = \! 30$~K and 
$T \! = \! 250$~K 
a smaller, slower decaying oscillatory signal is observed (see Figs.~\ref{fig3}(b)-(d)) indicating the 
occurrence of an additional kind of magnetic order in approximately 3~\% of the sample. 
The fact that we cannot follow this below $T \! = \! 30$~K is simply a consequence of the 8~\% 
impurity phase dominating the smaller signal. A substantially higher number of counts would be 
needed to simultaneously resolve both of these oscillatory components at low $T$. At $T \! = \! 30$~K 
the frequency of the 3~\% oscillatory signal is $\nu_2 \! = \! 1.92 \pm 0.15$~MHz, corresponding 
to a local field of $B_2 \! = \! 142 \pm 11$~G. Figures~\ref{fig3}(f)-(h) show the absence of these 
oscillatory signals in the $y \! = \! 6.57$ sample, which has the same hole-doping concentration as 
the large $y \! = \! 6.6$ single crystal. We have searched hard with higher counting statistics, and 
find no evidence for either oscillatory signal in any of the other YBCO samples or in a single
crystal of La$_{1.85}$Sr$_{0.15}$CuO$_4$ (see Fig.~\ref{fig3}(i)).

In a deliberate search for the CC phase in the $y \! = \! 6.6$ crystal with polarized neutron 
scattering \cite{Mook:08}, a transition to magnetic order that preserves the TSL was observed at 
$T \! = \! 235 \pm 15$~K. The results are similar to that reported earlier by Fauqu\'{e}
{\it et al.} \cite{Fauque:06}. The ordered moment is about 0.1~$\mu_B$ and is directed 
$55 \! \pm \! 7^{\circ}$ from the $c$-axis. To determine whether the 142~G field detected 
here by ZF-$\mu$SR is associated with the same magnetic order, we consider a model 
of staggered spins on the oxygen sites --- which was introduced in Ref.~\cite{Fauque:06} 
as one possible 
interpretation of the neutron measurements. Assuming the neutron scattering intensity associated 
with the magnetic order comes from only 3~\% of the sample, the true ordered moment is 
$(1/0.03)^{1/2} \! \times \! 0.1$~$\mu_B \! = \! 0.58$~$\mu_B$. A staggered moment of this size on the 
oxygen sites canted at $55 \! \pm \! 7^{\circ}$ (see Fig.~\ref{fig4}(c)) 
gives a resultant field of 
$107 \! \pm \! 5$~G and $192 \! \pm \! 14$~G at the chain and apical muon sites, respectively. 
The corresponding muon-site population average is $141 \! \pm \! 6$~G, which is in remarkable agreement 
with the average field of the 3~\% signal. Because the muon sites are outside the CuO$_2$ bilayers, 
reversing the direction of the spins in one of the CuO$_2$ planes of 
Fig.~\ref{fig4}(c) increases the field 
only slightly to $144 \! \pm \! 7$~G. Alternatively, circulating currents involving the apical 
oxygens \cite{Weber:09} give a muon-site-averaged field of about 127~G, which is also compatible 
with the observed precession frequency. However, the exceedingly small 3~\% 
magnetic volume fraction 
and the absence of a similar oscillating component in the other samples do 
not support the existence of a CC phase.  

While our experiments indicate that the unusual magnetic orders are associated with impurity phases,
we can only speculate on what these are. The first kind of magnetic order may be remnant traces of the 
starting material CuO. Bulk CuO exhibits AF phase transitions at $T_{N1} \! = \! 230$~K
and $T_{N2} \! = \! 213$~K, which are strongly coupled to the crystal lattice \cite{Yamada:04}.
A rise in magnetization also occurs in CuO near $T \! = \! 50$~K \cite{Okeefe:62}, where
lattice structure changes are observed in YBCO \cite{You:91,Islam:02}.
These different behaviors are noticeable in Fig.~\ref{fig1}, and as expected do not change 
appreciably with hole doping. We note that the onset of the ferromagnetic-like PKE signal in YBCO 
with $y \! = \! 6.92$ ($T_c \! = \! 92$~K) occurs at $T \! \sim \! 50$~K, and is
disproportionately weak compared to the PKE signal observed at higher temperatures in underdoped 
samples \cite{Xia:08}. As mentioned in Ref.~\cite{Xia:08}, impurities can induce a ferromagnetic 
component in an otherwise AF environment. At the very least our findings create ambiguity about 
the interpretation of the PKE signal near optimal doping, and hence 
whether there is a quantum critical point under
the superconducting ``dome''. 

Since variations in oxygen content are 
the doping mechanism for both YBCO and Hg1201, oxygen defects may play some role in the
second kind of unusual magnetic order. In YBCO,
short-range oxygen-vacancy ordering occurs in small dilute regions 
\cite{Islam:02,Strempfer:04,Islam:04}. However, something else more exotic, such as a small
concentration of Cu vacancies in the CuO$_2$ planes \cite{Elfimov:02} may be required to induce 
the observed magnetic order.   

We thank G.A.~Sawatzky for helpful discussions, and TRIUMF's 
Centre for Molecular and Materials Science for technical assistance. 
This work was supported by the Natural Sciences and Engineering Research Council of 
Canada and the Canadian Institute for Advanced Research.

\end{document}